# Product Constructions for Perfect Lee Codes

Tuvi Etzion, *Fellow, IEEE*

*Abstract*—A well known conjecture of Golomb and Welch is that the only nontrivial perfect codes in the Lee and Manhattan metrics have length two or minimum distance three. This problem and related topics were subject for extensive research in the last forty years. In this paper two product constructions for perfect Lee codes and diameter perfect Lee codes are presented. These constructions yield a large number of nonlinear perfect codes and nonlinear diameter perfect codes in the Lee and Manhattan metrics. A short survey and other related problems on perfect codes in the Lee and the Manhattan metrics are also discussed.

## I. INTRODUCTION

The Lee metric was introduced in [1], [2] for transmission of signals taken from GF($p$) over certain noisy channels. It was generalized for $\mathbb{Z}_m$ in [3]. The Lee distance $d_L(x,y)$ between two words $x = (x_1, x_2, \ldots, x_n)$, $y = (y_1, y_2, \ldots, y_n) \in \mathbb{Z}_m^n$ is given by $\Sigma_{i=1}^n \min\{x_i - y_i (\bmod\ m), y_i - x_i (\bmod\ m)\}$. A related metric, the Manhattan metric, is defined for alphabet letters taken as any integer. For two words $x = (x_1, x_2, \ldots, x_n)$, $y = (y_1, y_2, \ldots, y_n) \in \mathbb{Z}^n$ the Manhattan distance between $x$ and $y$ is defined as $d_M(x,y) \stackrel{\text{def}}{=} \Sigma_{i=1}^n |x_i - y_i|$. A code $\mathbb{C}$ in either metric (and in any other metric as well) has minimum distance $d$ if for each two distinct codewords $c_1, c_2 \in \mathbb{C}$ we have $d(c_1, c_2) \geq d$, where $d(\cdot, \cdot)$ stands for either the Lee distance or the Manhattan distance (or any other distance measure). Linear codes are usually the codes which can be handled more effectively and hence linear codes will be the building blocks in our constructions. We will not restrict ourself only for linear codes, but we will always assume that the all-zero word is a codeword.

A linear code in $\mathbb{Z}^n$ is an integer lattice. A *lattice* $\Lambda$ is a discrete, additive subgroup of the real $n$-space $\mathbb{R}^n$. W.l.o.g. (without loss of generality), we can assume that

$$\Lambda \stackrel{\text{def}}{=} \{u_1 v_1 + u_2 v_2 + \cdots + u_n v_n\ :\ u_1, u_2, \cdots, u_n \in \mathbb{Z}\} \quad (1)$$

where $\{v_1, v_2, \ldots, v_n\}$ is a set of linearly independent vectors in $\mathbb{R}^n$. A lattice $\Lambda$ defined by (1) is a sublattice of $\mathbb{Z}^n$ if and only if $\{v_1, v_2, \ldots, v_n\} \subset \mathbb{Z}^n$. We will be interested solely in sublattices of $\mathbb{Z}^n$. The vectors $v_1, v_2, \ldots, v_n$ are

T. Etzion is with the Department of Computer Science, Technion — Israel Institute of Technology, Haifa 32000, Israel. (email: etzion@cs.technion.ac.il).

This work was supported in part by the Israeli Science Foundation (ISF), Jerusalem, Israel, under Grant 230/08.

called *the basis* for $\Lambda \subseteq \mathbb{Z}^n$, and the $n \times n$ matrix

$$\mathbf{G} \stackrel{\text{def}}{=} \begin{bmatrix} v_{11} & v_{12} & \ldots & v_{1n} \\ v_{21} & v_{22} & \ldots & v_{2n} \\ \vdots & \vdots & \ddots & \vdots \\ v_{n1} & v_{n2} & \ldots & v_{nn} \end{bmatrix}$$

having these vectors as its rows is said to be the *generator matrix* for $\Lambda$. The lattice with the generator matrix $\mathbf{G}$ is denoted by $\Lambda(\mathbf{G})$.

*Remark 1:* There are other ways to describe a linear code in the Manhattan metric and the Lee metric. The traditional way of using a parity-check matrix can be also used [4], [5]. But, in our discussion, the lattice representation is the most convenient.

The *volume* of a lattice $\Lambda$, denoted $V(\Lambda)$, is inversely proportional to the number of lattice points per unit volume. More precisely, $V(\Lambda)$ may be defined as the volume of the *fundamental parallelogram* $\Pi(\Lambda)$, which is given by

$$\Pi(\Lambda) \stackrel{\text{def}}{=} \{\xi_1 v_1 + \xi_2 v_2 + \cdots + \xi_n v_n\ :\ 0 \leq \xi_i < 1,\ 1 \leq i \leq n\}$$

There is a simple expression for the volume of $\Lambda$, namely, $V(\Lambda) = |\det \mathbf{G}|$.

A shape $\mathcal{S}$ *tiles* $\mathbb{R}^n$ if disjoint copies of $\mathcal{S}$ cover all the points of $\mathbb{R}^n$. The cover of $\mathbb{R}^n$ with disjoint copies of $\mathcal{S}$ is called a *tiling*. We say that $\Lambda$ induces a *lattice tiling* of a shape $\mathcal{S}$ if disjoint copies of $\mathcal{S}$ placed on the lattice points on a given specific point in $\mathcal{S}$ form a tiling of $\mathbb{R}^n$.

Codes in $\mathbb{Z}^n$ generated by a lattice are periodic. We say that the code $\mathbb{C}$ has *period* $(m_1, m_2, \ldots, m_n) \in \mathbb{Z}^n$ if for each $i$, $1 \leq i \leq n$, the word $(x_1, x_2, \ldots, x_n) \in \mathbb{Z}^n$ is a codeword if and only if $(x_1, \ldots, x_{i-1}, x_i + m_i, x_{i+1}, \ldots, x_n) \in \mathbb{C}$. Let $m$ be the least common multiplier of the period $(m_1, m_2, \ldots, m_n)$. The code $\mathbb{C}$ has also period $(m, m, \ldots, m)$ and the code $\mathbb{C}$ can be reduced to a code $\mathbb{C}'$ in the Lee metric over the alphabet $\mathbb{Z}_m$ with the same minimum distance as $\mathbb{C}$. The parameters of such code will be given by $(n, d, v, m)$, where $n$ is the length of the code, $d$ is its minimum distance, $v$ is the volume of the related lattice ($\log_m v$ is the redundancy of the code), and $m$ is the alphabet size. The number of codewords in such a code is $\frac{m^n}{v}$.

The research on codes with the Manhattan metric is not extensive. It is mostly concern with the existence and nonexistence of perfect codes [3], [6], [7], [8]. Nevertheless, all codes defined in the Lee metric over some finite alphabet can be extended to codes in the Manhattan metric over the integers. The code resulting from a Lee code over $\mathbb{Z}_m$ is periodic with period $(m, m, \ldots, m)$. The minimum Manhattan distance will be the same as the minimum Lee distance. The literature on codes in the Lee metric is very extensive, e.g. [4], [9], [10], [11], [12], [13], [14], [15], [16]. The interest in Lee codes has been increased in the last decade due to many new and diverse applications of these codes. Some examples are constrained and and partial-response channels [4], interleaving schemes [17], multidimensional burst-error-correction [18], and error-correction for flash memories [19]. The increased interest is also due to new attempts to settle the existence question of perfect codes in these metrics [8].

Perfect codes is one of the most fascinating topics in coding theory. A perfect code in a given metric is a code in which the set of spheres with a given radius $R$ around its codewords form a partition of the space. These codes were mainly considered for the Hamming scheme, e.g. [20], [21], [22], [23], [24], [25], [26]. They were also considered for other schemes as the Johnson scheme and the Grassmann scheme, But, as said, these codes were considered to a larger extent also in the Lee and the Manhattan metrics.

This paper was motivated by some basic concepts which were presented in [27], [28]. There are two goals for this research. The first one is to present product constructions for perfect codes in the Lee and Manhattan metrics in a similar way to what was done in the Hamming scheme. The second one is to show how these constructions can be used to solve other problems related to perfect codes in the Lee and the Manhattan metrics. One problem is the number of different perfect codes and diameter perfect codes in the Lee and Manhattan metrics. A second problem is the existence of non-periodic perfect codes in the Manhattan metric. In the process, we will also present a new product construction for perfect codes in the Hamming scheme.

The rest of this paper is organized as follows. In Section II we introduce the basic concepts which are used in our paper. We first define a perfect code in general and discuss some of the known results on perfect codes and extended perfect codes in the Hamming scheme and the known results on perfect codes in the Lee and Manhattan metrics. We continue by introducing the concept of anticodes and diameter perfect code as was first discussed in [29]. We prove that the definition of diameter perfect codes can be applied to codes in the Lee and Manhattan metrics. We discuss the knowledge on these codes. We present a simple construction for diameter perfect codes with minimum distance four in the Lee and Manhattan metrics. Finally, we describe a doubling construction for perfect codes in the Hamming scheme which was given by Phelps in [20]. In Section III we present a new product construction for $q$-ary perfect codes in the Hamming scheme. Similar idea to the one in this construction was used in other papers, e.g. [24], [30], [31], [32], for construction of relatively large binary codes (perfect and non-perfect) with minimum Hamming distance three in the Hamming scheme. In Section IV we present two product constructions, one for perfect codes and one for diameter perfect codes in the Lee and Manhattan metrics. These constructions are modifications of the two product constructions in the Hamming scheme. In Section V we discuss two problems related to perfect codes and diameter perfect codes in the Lee and Manhattan metrics. The first one is the number of nonequivalent such codes and the second one is the existence of non-periodic perfect codes and non-periodic diameter perfect codes. We will show how our constructions can be used in the context of these two problems. A summary and a list of questions for future research are given in Section VI.

## II. BASIC CONCEPTS AND CONSTRUCTIONS

### A. Spheres and perfect codes

The main two concepts in this paper are perfect codes and codes in the Lee and the Manhattan metrics. We start with a general definition of perfect codes. For a given space $\mathcal{V}$, with a distance measure $d$, a subset $C$ of $\mathcal{V}$ is a *perfect code* with *radius* $R$ if for every element $x \in \mathcal{V}$ there exists a unique codeword $c \in C$ such that $d(x, c) \leq R$. For a point $x \in \mathcal{V}$, the *sphere* of radius $R$ around $x$, $S(x, R)$, is the set of elements in $\mathcal{V}$ such that $y \in S(x, R)$ if and only if $d(x, y) \leq R$, i.e. $S(x, R) = \{y \in \mathcal{V} \; : \; d(x, y) \leq R\}$. For the sphere $S(x, R)$, $x$ is called the *center* of the sphere. In this paper we consider only the Hamming scheme, the Lee and the Manhattan metrics, in which the size of a sphere does not depend on the center of the sphere. Hence, in the sequel we will assume that in our metric the size of a sphere with radius $R$ does not depend on its center. If $C$ is a code with minimum distance $2R+1$ and $S$ is a sphere with radius $R$ then it is readily verified that

*Theorem 1:* For a code $C$ with minimum distance $2R+1$ and a sphere $S$ with radius $R$ we have $|C| \cdot |S| \leq |\mathcal{V}|$.

Theorem 1 known as the *sphere packing bound*. In a code $C$ which attains the sphere packing bound, i.e. $|C| \cdot |S| = |\mathcal{V}|$, the spheres with radius $R$ around the codewords of $C$ form a partition of $\mathcal{V}$. Hence, such a code is a perfect code. A perfect code with Radius $R$ is also called a *perfect R-error-correcting code*.

Finally, a code $C$ of length $n$ which attains the sphere packing bound is determined by three parameters, its length, the radius of the sphere $R$ which is equivalent to the fact that $C$ has minimum distance $2R + 1$; the size of the code is $\frac{|\mathcal{V}|}{|S|}$, where $\mathcal{V}$ is the space on which the code is defined and $S$ is a sphere with radius $R$.

It is clear from the sphere packing bound that it is important to compute the size of a sphere in any given metric. We start with the size of a sphere in the Hamming scheme. For two words $x = (x_1, x_2, \ldots, x_n)$ and $y = (y_1, y_2, \ldots, y_n)$ over $\mathbb{Z}_m$ the *Hamming distance*, $d_H(x, y)$ is defined by

$$d_H(x, y) = |\{i \; : \; x_i \neq y_i, \; 1 \leq i \leq n\}| \; .$$

*Lemma 1:* The size of a sphere with radius $R$ centered in a word of length $n$ over $\mathbb{Z}_m$ in the Hamming scheme is

$$\sum_{i=0}^{R} \binom{n}{i}(m-1)^i \; .$$

*Corollary 1:* Let $\mathcal{C}$ be a perfect single-error-correcting code of length $n$ over an alphabet with $m$ letters in the Hamming scheme. Then the size of $\mathcal{C}$ is $\frac{m^n}{1+(m-1)n}$.

*Corollary 2:* If a code $\mathcal{C}$ of length $n$ over an alphabet with $m$ letters has minimum Hamming distance 3 and size $\frac{m^n}{1+(m-1)n}$ then $\mathcal{C}$ is a perfect code.

An *n-dimensional Lee sphere* with radius $R$, centered at $(\gamma_1, \gamma_2, \ldots, \gamma_n)$ is the shape $S_{n,R}$ in $\mathbb{Z}^n$ such that $(x_1, x_2, ..., x_n) \in S_{n,R}$ if and only if $\Sigma_{i=1}^n |x_i - \gamma_i| \leq R$, i.e., it consists of all points in $\mathbb{Z}^n$ whose Manhattan distance from the given point $(\gamma_1, \gamma_2, ..., \gamma_n) \in \mathbb{Z}^n$ is at most $R$. The size of $S_{n,R}$ is well known [3]:

$$|S_{n,R}| = \sum_{i=0}^{\min\{n,R\}} 2^i \binom{n}{i}\binom{R}{i} \qquad (2)$$

The $n$-dimensional Lee sphere with radius $R$ is the sphere with radius $R$ in the Manhattan metric and also in the Lee metric whenever the alphabet $\mathbb{Z}_m$ satisfies $m \geq 2R+1$.

*Corollary 3:* Let $\mathbb{C}$ be a perfect single-error-correcting code of length $n$ over $\mathbb{Z}_m$ in the Lee metric. Then the size of $\mathbb{C}$ is $\frac{m^n}{1+2n}$.

*Corollary 4:* If a code $\mathbb{C}$ of length $n$ over $\mathbb{Z}_m$ has minimum Lee distance 3 and size $\frac{m^n}{1+2n}$ then $\mathbb{C}$ is a perfect code.

If the code $\mathbb{C}$ is over $\mathbb{Z}$ and the code $\mathbb{C}$ is a perfect $R$-error-correcting code then the spheres with radius $R$ around the codewords form a tiling of $\mathbb{Z}^n$. Instead of Theorem 1 we have the following theorem.

*Theorem 2:* For a code $\mathbb{C} \subseteq \mathbb{Z}^n$ with minimum Manhattan distance $2R+1$, whose codewords are the points of a lattice $\Lambda$ we have $V(\Lambda) \geq |S_{n,R}|$.

### B. Perfect codes in the Hamming scheme

The existence question of perfect codes in the Hamming scheme was well investigated. It is well known [33] that the only parameters for nontrivial perfect codes over GF($q$) are those of the two Golay codes and the Hamming codes. Also over other alphabets it is highly probable that no new perfect codes exist [34]. The Hamming codes have length $\frac{q^r-1}{q-1}$, $r \geq 2$, where the alphabet is GF($q$). The minimum Hamming distance of the code is three. Codes with the parameters of the Hamming codes were extensively studied. A small sample of references includes [20], [21], [22], [24], [23], [25], [26].

An $[n,k,d]$ code $\mathcal{C}$ over $\mathbb{F}_q =$ GF($q$) is a linear subspace of dimension $k$ of $\mathbb{F}_q^n$. Clearly, $\mathcal{C}$ has $q^k$ codewords and $q^{n-k}$ cosets. For any code $\mathcal{C}$ (linear or nonlinear) of length $n$ over $F_q =$ GF($q$) and a word $v \in F_q^n$ we define a *translate* of $\mathcal{C}$ with the word $v$ by

$$v + \mathcal{C} \stackrel{\text{def}}{=} \{v + c \ : \ c \in \mathcal{C}\} .$$

If $\mathcal{C}$ is a linear code then each translate is a *coset* of the code. It is easy to verify that the size of a translate $v + \mathcal{C}$ is equal to the size of $\mathcal{C}$. If $\mathcal{C}$ is linear and $v_1, v_2 \in \mathbb{F}_q^n$ then either $v_1 + \mathcal{C} = v_2 + \mathcal{C}$ or $(v_1 + \mathcal{C}) \cap (v_2 + \mathcal{C}) = \varnothing$.

Therefore, the cosets of the code form a partition of $\mathbb{F}_q^n$. If $\mathcal{C}$ is nonlinear then this is usually not true. But, if $\mathcal{C}$ is a perfect code then some translates form a partition of $\mathbb{F}_q^n$. For example, if $\mathcal{C}$ is a perfect single-error-correcting code and $e_i$ and $e_j$ are two different unit vectors of length $n$ then $(e_i + \mathcal{C}) \cap (e_j + \mathcal{C}) = \varnothing$. Therefore, the code $\mathcal{C}$ and all its translates formed from the unit vectors form a partition of $\mathbb{F}_q^n$.

Given a binary perfect single-error-correcting code $\mathcal{C}$ of length $n$, one can define the extended code $\mathcal{C}_e$ by

$$\mathcal{C}_e \stackrel{\text{def}}{=} \{(c, p(c)) \ : \ c \in \mathcal{C}, \ p(c) = \sum_{i=1}^n c_1 \ (mod \ 2)\} ,$$

where $p(c)$ is the *parity* of $c$, i.e., *zero* if the weight of $c$ (number of ones in $c$) is even and *one* if the weight of $c$ is odd. The extended code $\mathcal{C}_e$ has length $n+1$, size exactly as the size of $\mathcal{C}$, and minimum Hamming distance four. All the codewords of $\mathcal{C}_e$ have even weight. The extended code $\mathcal{C}_e$ has $n+1$ odd translates (all words have odd weight) consisting of all translates from unit vectors. It has $n+1$ even translates (all words have even weight) consisting of the code $\mathcal{C}_e$ and all translates from vectors of weight two with an *one* in the last coordinate. All these $2n+2$ translates are disjoint and together form a partition of $\mathbb{F}_2^{n+1}$. A binary perfect single-error-correcting code $\mathcal{C}$ has length $2^r - 1$ and $2^{n-r}$ codewords. Let $\mathcal{C}'$ be a code of length $2^r$, minimum distance four, and $2^{n-r}$ codewords all with even weight. It is easy to verify that the punctured code

$$\mathcal{C}^p \stackrel{\text{def}}{=} \{c \ : \ (c, p(c)) \in \mathcal{C}'\}$$

has length $2^r - 1$, minimum distance three, and $2^{n-r}$ codewords. Hence, by Corollary 2, $\mathcal{C}^p$ is a perfect code.

### C. Perfect codes in the Lee and Manhattan metrics

The existence question of perfect $R$-error-correcting codes in the Lee metric was first asked by Golomb and Welch in their seminal paper [3]. They constructed perfect codes for length $n = 2$ over $\mathbb{Z}_m$, where $m = 2R^2 + 2R + 1$, and the minimum Lee distance of the code is $2R+1$. They also constructed perfect single-error-correcting codes of length $n$ over $\mathbb{Z}_m$, $m = 2n+1$. They also considered perfect codes in the Manhattan metric. They proved that for each $n > 3$ there a exists an integer $\rho_n$ such that for each radius $R > \rho_n$ there is no tiling of the $n$-dimensional Euclidian space with the sphere $S_{n,R}$, i.e., there is no length $n$, perfect $R$-error-correcting code in the Manhattan metric. The existence problem of other perfect codes was discussed in many papers, e.g. [3], [8], [9], [10], [12]. We will mention two of the results given in these papers. Post [9] proved that if $m \geq 2R+1$ then there are no perfect Lee codes over $\mathbb{Z}_m$ for $3 \leq n \leq 5$, $R \geq n-2$, and for $n \geq 6$, $R \geq \frac{\sqrt{2}}{2}n - \frac{1}{4}(3\sqrt{2} - 2)$. Another interesting result was given in [12] where it was proved that the smallest $m$ for which there exists a perfect single-error-correcting Lee code over $\mathbb{Z}_m$ is the multiplication of all the prime factors of $2n+1$.

In the same way as was done for the Hamming scheme, a perfect single-error-correcting code $\mathbb{C}$ (in the Lee metric) and its translates formed from the unit vectors (vectors of length $n$ with exactly one nonzero entry having value -1 or +1) form a partition of $\mathbb{Z}_m^n$. If the perfect code is over $\mathbb{Z}$ then the partition is of $\mathbb{Z}^n$. In contrary to the Hamming scheme, not many constructions are known for perfect single-error-correcting codes in the Lee metric. Not many properties are known, and there is no "reasonable" lower bound on the number of nonequivalent perfect single-error-correcting codes.

Finally, if $m < 2R+1$ then $S_{n,R}$ is not the sphere in the corresponding parameters of the Lee metric. If $m = 2$ or $m = 3$ then the Lee metric coincides with the Hamming scheme for binary codes or ternary codes, respectively. If $m > 3$ then the situation becomes more interesting, but also more complicated.

### D. Anticodes and diameter perfect codes

In all the perfect codes the minimum distance of the code is an odd integer. If the minimum distance of the code $C$ is an even integer then there cannot be any perfect code since for any two codewords $c_1, c_2 \in C$ such that $d(c_1, c_2) = 2\delta$ there exists a word $x$ such that $d(x, c_1) = \delta$ and $d(x, c_2) = \delta$. For this case another concept is used, a diameter perfect code, as was defined in [29]. This concept is based on the code-anticode bound presented by Delsarte [35]. An *anticode* $\mathcal{A}$ of *diameter* $D$ in a space $\mathcal{V}$ is a subset of words from $\mathcal{V}$ such that $d(x, y) \leq D$ for all $x, y \in \mathcal{A}$.

*Theorem 3:* If a code $C$, in a space $\mathcal{V}$ of a distance regular graph, has minimum distance $d$ and in the anticode $\mathcal{A}$ of the space $\mathcal{V}$ the maximum distance is $d-1$ then $|C| \cdot |\mathcal{A}| \leq |\mathcal{V}|$.

Theorem 3 which is proved in [35] is applied to the Hamming scheme since the related graph is distance regular. It cannot be applied to the Manhattan metric since the related graph is not finite. It cannot be applied to the Lee metric since the related graph is not distance regular. This can be easily verified by considering the three words $x = (000\ldots0)$, $y = (200\ldots0)$, and $z = (110\ldots0)$ of length $n$ over $\mathbb{Z}_m$, $m \geq 5$. $d_L(x,y) = d_L(x,z) = 2$; there exists exactly one word $u$ for which $d_L(x, u) = 1$ and $d_L(u, y) = 1$, while there are exactly two words of the form $u$ for which $d_L(x, u) = 1$ and $d_L(u, z) = 1$. Fortunately, an alternative proof which was given in [29] can be slightly modified to work for the Lee metric.

*Theorem 4:* Let $\mathbb{C}_\mathcal{D}$ be a code of length $n$ over $\mathbb{Z}_m$ with Lee distances between codewords taken from a set $\mathcal{D}$. Let $\mathcal{A} \subset \mathbb{Z}_m^n$ and let $\mathbb{C}'_\mathcal{D}$ be the largest code in $\mathcal{A}$ with Lee distances between codewords taken from the set $\mathcal{D}$. Then

$$\frac{|\mathbb{C}_\mathcal{D}|}{m^n} \leq \frac{|\mathbb{C}'_\mathcal{D}|}{|\mathcal{A}|} .$$

*Proof:* Let $S \stackrel{\text{def}}{=} \{(c, v) : c \in \mathbb{C}_\mathcal{D}, v \in \mathbb{Z}_m^n, c+v \in \mathcal{A}\}$. For a given codeword $c \in \mathbb{C}_\mathcal{D}$ and a word $\alpha \in \mathcal{A}$, there is exactly one element $v = c + \alpha \in \mathbb{Z}_m^n$ such that $\alpha = c + v$. Therefore, $|S| = |\mathbb{C}_\mathcal{D}| \cdot |\mathcal{A}|$.

Since $\mathbb{C}'_\mathcal{D}$ is the largest code in $\mathcal{A}$, with Lee distances between codewords taken from the set $\mathcal{D}$, it follows that for any given word $v \in \mathbb{Z}_m^n$ the set $\{c : c \in \mathbb{C}_\mathcal{D}, c+v \in \mathcal{A}\}$ has at most $|\mathbb{C}'_\mathcal{D}|$ codewords. Hence, $|S| \leq |\mathbb{C}'_\mathcal{D}| \cdot m^n$.

Thus, $|\mathbb{C}_\mathcal{D}| \cdot |\mathcal{A}| \leq |\mathbb{C}'_\mathcal{D}| \cdot m^n$ and the claim is proved. ∎

*Corollary 5:* Theorem 3 holds for the Lee metric, i.e. if a code $\mathbb{C} \subset \mathbb{Z}_m^n$, has minimum Lee distance $d$ and in the anticode $\mathcal{A} \subset \mathbb{Z}_m^n$ the maximum Lee distance is $d-1$ then $|\mathbb{C}| \cdot |\mathcal{A}| \leq m^n$.

*Proof:* Let $\mathcal{D} = \{d, d+1, \ldots, \lfloor \frac{m}{2} \rfloor n\}$ and let $\mathbb{C}_\mathcal{D}$ be a code from $\mathbb{Z}_m^n$ with minimum Lee distance $d$. Let $\mathcal{A}$ be a subset of $\mathbb{Z}_m^n$ with Lee distances between words of $\mathcal{A}$ taken from the set $\{1, 2, \ldots, d-1\}$. i.e. $\mathcal{A}$ is an anticode with diameter $d-1$. Clearly, the largest code in $\mathcal{A}$ with Lee distances from $\mathcal{D}$ has only one codeword. Applying Theorem 4 on $\mathcal{D}$, $\mathbb{C}_\mathcal{D}$, and $\mathcal{A}$, implies $\frac{|\mathbb{C}_\mathcal{D}|}{m^n} \leq \frac{1}{|\mathcal{A}|}$, i.e. $|\mathbb{C}_\mathcal{D}| \cdot |\mathcal{A}| \leq m^n$.

Thus, the claim is proved. ∎

Corollary 5 can be stated similarly to Theorem 2.

*Theorem 5:* Let $\Lambda$ be a lattice which forms a code $\mathbb{C} \subset \mathbb{Z}^n$ with minimum Manhattan distance $d$. Then the size of any anticode of length $n$ with maximum distance $d-1$ is at most $V(\Lambda)$.

A code which attains either the bounds of Theorem 3 or the bound of Corollary 5 or the bound of Theorem 5 with equality is called a *diameter perfect code*. The definition in a space which consists of the vertices from a distance regular graph was first given by Ahlswede, Aydinian, and Khachatrian [29].

What is the size of the largest anticode with Diameter $D$ in $\mathbb{Z}^n$, $n \geq 2$? The answer was given in [36, pp. 30–41]. If $D = 2R$ then this anticode is an $n$-dimensional Lee sphere with radius $R$, $S_{n,R}$. For odd $D$ we define anticodes with diameter $D = 2R+1$ as follows. For $R = 0$, let $S'_{n,0}$ be a shape which consists of two adjacent points of $\mathbb{Z}^n$, where two points are adjacent if the Manhattan distance between them is one. $S'_{n,R+1}$ is defined by adding to $S'_{n,R}$ all the points which are adjacent to at least one of its points. $S'_{n,R}$ is the largest anticode in $\mathbb{Z}^n$ with diameter $2R+1$. $S_{2,R}$ and $S'_{2,R}$ were used in [17] for two-dimensional interleaving schemes to correct two-dimensional cluster errors. Similarly to the computation of (2) in [3] one can compute the size of $S'_{n,R}$ [27] and obtain

$$|S'_{n,R}| = \sum_{i=0}^{\min\{n-1, R\}} 2^{i+1} \binom{n-1}{i} \binom{R+1}{i+1} . \quad (3)$$

Ahlswede, Aydinian, and Khachatrian [29] have examined the Hamming, Johnson and Grassmann schemes for the existence of diameter perfect codes. All perfect codes in these schemes are also diameter perfect codes. In the Hamming schemes there are two more families of diameter perfect codes. MDS codes [33] are diameter perfect codes, extended Golay codes, and extended binary perfect single-error-correcting codes are also diameter perfect codes.

Finally, similarly to Corollary 2 and Corollary 4 we have the following result.

*Lemma 2:* If a code $\mathbb{C}$ of length $n$ over $\mathbb{Z}_m$ has minimum Lee distance 4 and size $\frac{m^n}{4n}$ then $\mathbb{C}$ is a diameter perfect code.

### E. Diameter perfect codes in the Lee metric

By (3) the size of a maximum anticode of length $n$ with diameter three over $\mathbb{Z}^n$ in the Manhattan metric is $4n$. Therefore, by Theorem 5, a related diameter perfect code with minimum distance four can be formed from a lattice with volume $4n$ in which the minimum Manhattan distance is four. Consider the lattice $\Lambda(G_n)$ defined in [27] by the following generator matrix

$$G_n = \left[ \begin{array}{cc} A_n & B_n \\ C_n & D_n \end{array} \right],$$

where $A_n = I_{n-1}$, $B_n$ is the $(n-1) \times 1$ matrix for which $B_n^T = [3\ 5\ \cdots\ 2n-1]$, $C_n$ is an $1 \times (n-1)$ all-zero matrix, and $D_n = [4n]$.

*Example 1:* For $n = 6$, $G_6$ has the form

$$G_6 = \left[ \begin{array}{cccccc} 1 & 0 & 0 & 0 & 0 & 3 \\ 0 & 1 & 0 & 0 & 0 & 5 \\ 0 & 0 & 1 & 0 & 0 & 7 \\ 0 & 0 & 0 & 1 & 0 & 9 \\ 0 & 0 & 0 & 0 & 1 & 11 \\ 0 & 0 & 0 & 0 & 0 & 24 \end{array} \right]$$

The volume of $\Lambda(G_n)$ is $4n$ and it is easy to verify that the minimum Manhattan distance of the code defined by $\Lambda(G_n)$ is four. Moreover, it is readily verified that $\Lambda(G_n)$ is reduced to a code over $\mathbb{Z}_{4n}$. Hence, we have the following theorem.

*Theorem 6:* The code defined by the lattice $\Lambda(G_n)$ is a diameter perfect code, with minimum distance four, in the Manhattan metric. The code can be reduced to a Lee code $\mathbb{C}$ over $\mathbb{Z}_{4n}$. $\mathbb{C}$ is a diameter perfect code, with minimum distance four, in the Lee metric.

The code defined by $\Lambda(G_n)$ is not unique. There are other diameter perfect codes of length $n$ and minimum Lee distance four. Some of these codes will be considered in Section IV. The main goal will be to construct many such codes over alphabets with the smallest possible size.

For $n = 2$ there are diameter perfect codes for each even minimum Lee distance. By (3) the size of a maximum anticode of length 2 with diameter $2R + 1$ over $\mathbb{Z}$ is $2(R+1)^2$. For each $i$, $0 \leq i \leq R$ the following generator matrix forms a lattice which generates a diameter perfect code with minimum distance $2R$.

$$\left[ \begin{array}{cc} R+1+i & R+1-i \\ i & 2(R+1)-i \end{array} \right].$$

These lattices will be interleaved together to form more diameter perfect codes and non-periodic diameter perfect codes in Section V.

Cosets and translates of a code $\mathbb{C}$ are defined in the Lee and Manhattan metrics in the same way that they are defined in the Hamming scheme. The *Lee (Manhattan) weight* of a word $x = (x_1, x_2, \ldots, x_n)$ over $\mathbb{Z}_m$ ($\mathbb{Z}$) is defined by $d_L(x, \mathbf{0})$ ($d_M(x, \mathbf{0})$), where $\mathbf{0}$ is the all-zero word. A translate of a code $\mathbb{C}$ is said to be an even translate if all its words have even Lee (Manhattan) weight. A translate is said to be an odd translate if all its words have odd Lee (Manhattan) weight.

Similarly to to extended perfect single-error-correcting codes in the Hamming scheme we have even translates and odd translates for diameter perfect codes with even distance. This is proved in the following results. The first lemma can be readily verified.

*Lemma 3:* A diameter perfect code $\mathbb{C}$ of length $n$, over $\mathbb{Z}_m$, with minimum Lee distance $d$ can be extended to a diameter perfect code $\mathbb{C}'$ of length $n$, over $\mathbb{Z}$, with minimum Manhattan distance $d$.

*Lemma 4:* If $\mathbb{C}$ is a diameter perfect code with minimum Manhattan distance $d$, $d$ even, then all the codewords of $\mathbb{C}$ have even Manhattan weight.

*Proof:* Let $\mathbb{C}$ be a diameter perfect code with minimum distance $2R + 2$, in the Manhattan metric. Assume the contrary that there exists a codeword with odd Manhattan weight. Since $\mathbb{C}$ is a perfect diameter code it follows that there exists a tiling of $\mathbb{R}^n$ with the anticode $S'_{n,R}$ whose diameter is $2R + 1$. W.l.o.g. we can assume that the two adjacent points in $S'_{n,0}$ (the core of each $S'_{n,R}$ in this tiling) differ in the last coordinate, one of these points $(z_1, \ldots, z_{n-1}, z_n)$ is a codeword in $\mathbb{C}$ and the second point is $(z_1, \ldots, z_{n-1}, z_n + 1)$. In the tiling there exists two points $x = (x_1, \ldots x_{n-1}, x_n), y = (y_1, \ldots y_{n-1}, y_n) \in \mathbb{Z}^n$ such that $d_M(x, y) = 1$ for which $x$ is in an anticode which contains the codeword $\alpha \in \mathbb{C}$ and $y$ is in an anticode which contains the codeword $\beta \in \mathbb{C}$; and furthermore, $\alpha$ has even Manhattan weight and $\beta$ has odd Manhattan weight. Clearly, $d_M(x, \alpha), d_M(y, \beta) \in \{R, R+1\}$, $d_M(\alpha, \beta)$ is odd and at least $2R + 2$. It implies that $d_M(x, \alpha) = d_M(y, \beta) = R + 1$ and $d_M(\alpha, \beta) = 2R + 3$. $d_M(x, \alpha) = d_M(y, \beta) = R + 1$ and the facts that in the tiling $x$ is contained in the anticode of $\alpha$ and $y$ is contained in the anticode of $\beta$ implies that $x_n$ is greater than the last entry of $\alpha$ and $y_n$ is greater than the last entry of $\beta$. It follows that $d_M((x_1, \ldots, x_{n-1}, x_n - 1), \alpha) = d_M((y_1, \ldots, y_{n-1}, y_n - 1), \beta) = R$ and since $d_M(x, y) = 1$ we have that $d_M(\alpha, \beta) \leq 2R + 1$, a contradiction.

Thus, if $\mathbb{C}$ is a diameter perfect code with minimum Manhattan distance $d$, $d$ even, then all the codewords of $\mathbb{C}$ have even Manhattan weight. ∎

*Lemma 5:* If $\mathbb{C}$ is a diameter perfect code with minimum Lee distance $d$, $d$ even, over $\mathbb{Z}_m$, then $m$ is even.

*Proof:* By (3) the size of the anticode with maximum distance $d - 1$ is even. This implies by the definition of a diameter perfect code that the size of the space is even. Thus, $m$ is even. ∎

*Corollary 6:* If $\mathbb{C}$ is a diameter perfect code with minimum Lee distance $d$, $d$ even, then all the codewords of $\mathbb{C}$ have even Lee weight.

*Theorem 7:* Each translate of a diameter perfect code with even distance, in the Lee metric, is either an even translate or an odd translate. The number of even translates is equal the number of odd translates.

*Proof:* Let $\mathbb{C}$ be a diameter perfect code in the Lee metric. By Lemma 5 and Corollary 6 each translate of $\mathbb{C}$ is either even translate or an odd translate. W.l.o.g. we can assume that the two points of $S'_{n,0}$ in $S'_{n,R}$ are $\alpha = (0,\ldots,0,0)$ and $\beta = (0,\ldots,0,1)$. Let $\mathbb{C}$ a diameter perfect code of length $n$ and diameter $2R+1$, in the Lee metric, where $\alpha \in \mathbb{C}$. Since $S'_{n,R}$ is symmetric around $\alpha$ and $\beta$ we have that for each $0 \leq \delta \leq R+1$, $|\{x : x \in S'_{n,R}, d_L(x,\alpha) = \delta\}| = |\{x : x \in S'_{n,R}, d_L(x,\beta) = \delta\}|$, and for each point $x$, $d_L(x,\alpha) = d_L(x,\beta) + 1 \pmod{2}$. Therefore, the number of even translates of $\mathbb{C}$ is equal the number of odd translates of $\mathbb{C}$. ∎

We note on one important difference between binary perfect codes and binary diameter perfect codes in the Hamming scheme and perfect codes and diameter perfect codes in the Lee metric. While binary perfect single-error-correcting codes in the Hamming scheme can be extended to binary diameter perfect codes (and vice versa via puncturing) by adding a parity bit (removing a coordinate) and thus increasing (decreasing) the distance by one, this cannot be done in the Lee metric.

Finally, there is one more diameter perfect code in the Lee and Manhattan metric. It is formed by the Minkowski lattice [37] with the generator matrix

$$\begin{bmatrix} 1 & -2 & 3 \\ -2 & 3 & 1 \\ 3 & 1 & -2 \end{bmatrix}.$$

The related anticode in this case is $S'_{3,2}$.

### F. A doubling construction for binary perfect codes

Let $\mathcal{C}$ be a binary perfect single-error-correcting code of length $n = 2^r - 1$ and let $\mathcal{C}_e$ its extended code of length $2^r$. Let $\mathcal{C}^1 = \mathcal{C}, \mathcal{C}^2, \ldots, \mathcal{C}^{2^r}$ be its $2^r$ translates and $\mathcal{C}_e^1 = \mathcal{C}_e, \mathcal{C}_e^2, \ldots, \mathcal{C}_e^{2^r}$ be the $2^r$ even translates of its extended code. Let $\mathcal{C}^*$ be the code constructed as follows:

$$\mathcal{C}^* \stackrel{\text{def}}{=} \{(x,y) : x \in \mathcal{C}_e^i, y \in \mathcal{C}^i, 1 \leq i \leq 2^r\}$$

It was proved by Phelps [20] that

*Theorem 8:* $\mathcal{C}^*$ is a perfect single-error-correcting code of length $2^{r+1} - 1$.

It is readily verified that the perfect extended code $\mathcal{C}_e^*$ is the code defined by

$$\{(x,y) : x \in \mathcal{C}_e^i, y \in \mathcal{C}_e^i, 1 \leq i \leq 2^r\}$$

Let $\mathcal{B}$ be a binary perfect single-error-correcting code of length $n = 2^r - 1$ and let $\mathcal{B}_e$ its extended code of length $2^r$. Let $\mathcal{B}_e^1 = \mathcal{B}_e, \mathcal{B}_e^2, \ldots, \mathcal{B}_e^{2^r}$ be the $2^r$ even translates of its extended code. Let $\pi = (\pi(1) = 1, \pi(2), \ldots, \pi(2^r))$ be a permutation of $\{1, 2, \ldots, 2^r\}$. The following theorem is an immediate consequence

*Theorem 9:* The code defined by

$$\mathcal{C}^* \stackrel{\text{def}}{=} \{(x,y) : x \in \mathcal{B}_e^i, y \in \mathcal{C}_e^{\pi(i)}, 1 \leq i \leq 2^r\}$$

is an extended perfect code of length $2^{r+1}$ with minimum Hamming distance four.

Note, that the first element in the permutation $\pi$ is 1 to make sure that the all-zero word will be a codeword in $\mathcal{C}^*$.

## III. A NEW PRODUCT CONSTRUCTION FOR $q$-ARY PERFECT CODES

There are several product constructions for non-binary perfect codes in the Hamming scheme. Most notable is the general construction of Phelps [22]. Another construction was given by Mollard [23]. The construction that we present is a generalization for a construction of Zinov'ev [24].

We will present a new simple construction, for perfect codes in the Hamming scheme, which will be very effective in constructions of perfect codes in the Lee metric. For our construction we will use two perfect codes in the Hamming scheme. The first code $\mathcal{C}_1$ is a perfect single-error-correcting code of length $n = \frac{q^r-1}{q-1}$ over an alphabet with $q$ letters, which have a total of $q^r$ translates, including $\mathcal{C}_1$ itself. The second code $\mathcal{C}_2$ is a perfect single-error-correcting code of length $\ell = \frac{q^{rs}-1}{q^r-1}$ over an alphabet with $q^r$ letters. Let $\mathcal{C}_1^i$, $1 \leq i \leq q^r$, be the $i$th translate of $\mathcal{C}_1$, where $\mathcal{C}_1^1 = \mathcal{C}_1$. We construct the following code $\mathcal{C}^*$:

$$\mathcal{C}^* \stackrel{\text{def}}{=} \{(x_{i_1}, x_{i_2}, \ldots, x_{i_\ell}) : x_{i_t} \in \mathcal{C}_1^{i_t}, (i_1, i_2, \ldots, i_\ell) \in \mathcal{C}_2\}$$

*Theorem 10:* The code $\mathcal{C}^*$ is a $q$-ary perfect single-error-correcting code of length $\frac{q^{rs}-1}{q-1}$.

*Proof:* Clearly, the length of the codewords from $\mathcal{C}^*$ is $\frac{q^r-1}{q-1} \frac{q^{rs}-1}{q^r-1} = \frac{q^{rs}-1}{q-1}$. We will first prove that the minimum distance of $\mathcal{C}^*$ is three. Let $(x_{i_1}, x_{i_2}, \ldots, x_{i_\ell})$ and $(y_{j_1}, y_{j_2}, \ldots, y_{j_\ell})$ be two distinct codewords of $\mathcal{C}^*$. We now distinguish between two cases.

**Case 1:** If $(i_1, i_2, \ldots, i_\ell) \neq (j_1, j_2, \ldots, j_\ell)$ then $d_H((i_1, i_2, \ldots, i_\ell), (j_1, j_2, \ldots, j_\ell)) \geq 3$ since $(x_{i_1}, x_{i_2}, \ldots, x_{i_\ell}), (y_{j_1}, y_{j_2}, \ldots, y_{j_\ell}) \in \mathcal{C}_2$ and $d_H(\mathcal{C}_2) = 3$. W.l.o.g., we can assume that $i_1 \neq j_1$, $i_2 \neq j_2$, and $i_3 \neq j_3$, and hence $x_{i_1} \neq y_{j_1}$, $x_{i_2} \neq y_{j_2}$, and $x_{i_3} \neq y_{j_3}$, which implies that $d_H((x_{i_1}, x_{i_2}, \ldots, x_{i_\ell}), (y_{j_1}, y_{j_2}, \ldots, y_{j_\ell})) \geq 3$.

**Case 2:** If $(i_1, i_2, \ldots, i_\ell) = (j_1, j_2, \ldots, j_\ell)$ then there exists a $t$, $1 \leq t \leq \ell$, such that $x_{i_t} \neq y_{j_t}$ and since $i_t = j_t$ (i.e. $\mathcal{C}_1^{i_t} = \mathcal{C}_1^{j_t}$), $x_{i_t}, y_{j_t} \in \mathcal{C}_1^{i_t}$, and $d_H(\mathcal{C}_1^{i_t}) = 3$, it follows that $d(x_{i_t}, y_{j_t}) \geq 3$ and therefore $d((x_{i_1}, x_{i_2}, \ldots, x_{i_\ell}), (y_{j_1}, y_{j_2}, \ldots, y_{j_\ell})) \geq 3$.

Thus, $d_H(\mathcal{C}^*) \geq 3$.

By Corollary 1, the size of $\mathcal{C}_1$ is $\frac{q^n}{1+(q-1)n} = q^{n-r}$ and the size of $\mathcal{C}_2$ is $\frac{q^{r\ell}}{1+(q^r-1)\ell} = q^{r\ell-rs}$. Clearly, $|\mathcal{C}^*| = |\mathcal{C}_2| \cdot |\mathcal{C}_1|^\ell = q^{r\ell-rs}q^{(n-r)\ell} = q^{n\ell-rs} = \frac{q^{n\ell}}{1+(q-1)n\ell}$.

This implies by Corollary 2 that $\mathcal{C}^*$ is a $q$-ary perfect single-error-correcting code of length $\frac{q^{rs}-1}{q-1}$. ∎

## IV. CONSTRUCTIONS FOR PERFECT AND DIAMETER PERFECT LEE CODES

In this section we will modify the two product constructions for perfect codes and diameter perfect codes in the Hamming scheme to obtain perfect codes and diameter perfect codes in the Lee and the Manhattan metrics.

### A. Diameter perfect codes with minimum distance four

Let $\mathbb{C}_1$ and $\mathbb{C}_2$ be $(n, 4, 4n, m)$ diameter perfect codes. Each code has $4n$ translates from which $2n$ are even translates. Let $\mathbb{C}_r^1 = \mathbb{C}_r, \mathbb{C}_r^2, \ldots, \mathbb{C}_r^{2n}$, $r = 1, 2$, be these $2n$ even translates. Let $\pi = (\pi(1) = 1, \pi(2), \ldots, \pi(2n))$ be a permutation of $\{1, 2, \ldots, 2n\}$. The following theorem is a modification of Theorem 9.

*Theorem 11:* The code $\mathbb{C}^*$ defined by

$$\mathbb{C}^* \stackrel{\text{def}}{=} \{(x, y) \ : \ x \in \mathbb{C}_1^i, \ y \in \mathbb{C}_2^{\pi(i)}, \ 1 \leq i \leq 2n\}$$

is an diameter perfect Lee code of length $2n$, over $\mathbb{Z}_m$, with minimum Lee distance four.

*Proof:* The size of the code $\mathbb{C}_r$, $r = 1, 2$, is $\frac{m^n}{4n}$ and hence the size of $\mathbb{C}^*$ is $2n\frac{m^{2n}}{16n^2} = \frac{m^{2n}}{8n}$. The Lee distance of the code is four as the one given in Theorem 9. Since no proof is given in subsection II-F we will now present a proof. Let $(x_1, x_2)$ and $(y_1, y_2)$ be two distinct codewords in $\mathbb{C}^*$ such that $x_1 \in \mathbb{C}_1^i$ and $y_1 \in \mathbb{C}_1^j$. We distinguish between two cases.

**Case 1:** If $i \neq j$ then $\mathbb{C}_1^i \neq \mathbb{C}_1^j$ and $\mathbb{C}_2^{\pi(i)} \neq \mathbb{C}_2^{\pi(j)}$. Hence $d_L(x_1, y_1) \geq 2$, $d_L(x_2, y_2) \geq 2$, which implies that $d_L((x_1, x_2), (y_1, y_2)) \geq 4$.

**Case 2:** If $i = j$ then $x_1, y_1 \in \mathbb{C}_1^i$ and $x_2, y_2 \in \mathbb{C}_2^{\pi(i)}$. Since $x_1 \neq y_1$ or $x_2 \neq y_2$ we have that $d_L(x_1, y_1) \geq 4$ or $d_L(x_2, y_2) \geq 4$, respectively. Hence, $d_L((x_1, x_2), (y_1, y_2)) \geq 4$.

Thus $d_L(\mathbb{C}^*) \geq 4$.

The code $\mathbb{C}^*$ is defined over $\mathbb{Z}_m$ and hence the size of its space is $m^{2n}$. Since by (3) the size of the related anticode with diameter four is $8n$ and the minimum Lee distance of $\mathbb{C}^*$ is four, it follows by Lemma 2 that $\mathbb{C}^*$ is a diameter perfect Lee code of length $2n$, over $\mathbb{Z}_m$, with minimum Lee distance four. ■

*Corollary 7:* Let $n = 2^r p$, where $p$ is an odd prime greater than one. There exists an $(n, 4, 4n, 4p)$ diameter perfect code.

The lattice of the following generator matrix

$$\begin{bmatrix} 2 & 2 \\ 0 & 4 \end{bmatrix},$$

form a $(2, 4, 8, 4)$ diameter perfect code. By applying Theorem 11 iteratively we obtain the following theorem.

*Theorem 12:* For each $n = 2^r$, $r \geq 1$, there exists an $(n, 4, 4n, 4)$ diameter perfect code.

Codes with the same parameters as in Theorem 12 were generated by Krotov [38].

Theorem 11 can be modified and applied on codes in $\mathbb{Z}^n$ with the Manhattan metric.

*Theorem 13:* Let $\mathbb{C}_1$ and $\mathbb{C}_2$ be diameter perfect codes of length $n$ with minimum Manhattan distance four. Each code has $4n$ translates from which $2n$ are even translates. Let $\mathbb{C}_i^1 = \mathbb{C}_i, \mathbb{C}_i^2, \ldots, \mathbb{C}_i^{2n}$, $i = 1, 2$, be these $2n$ even translates. Let $\pi = (\pi(1) = 1, \pi(2), \ldots, \pi(2n))$ be a permutation of $\{1, 2, \ldots, 2n\}$. The code $\mathbb{C}^*$ defined by

$$\mathbb{C}^* \stackrel{\text{def}}{=} \{(x, y) \ : \ x \in \mathbb{C}_1^i, \ y \in \mathbb{C}_2^{\pi(i)}, \ 1 \leq i \leq 2n\} \quad (4)$$

is a diameter perfect Manhattan code of length $2n$, over $\mathbb{Z}$, with minimum Manhattan distance four.

### B. Perfect single-error-correcting Lee codes

In this subsection we use a modification of the product construction of subsection III to construct perfect single-error-correcting Lee codes.

Let $\mathbb{C}_1$ be a perfect single-error-correcting Lee code of length $n = \frac{q^r - 1}{2}$, $q$ odd, over an alphabet with $\tau(2n+1)$ letters, which has a total of $q^r$ translates (the size of the Lee sphere), including $\mathbb{C}_1$ itself. Let $\pi_t = (\pi_t(1) = 1, \pi_t(2), \ldots, \pi_t(2n))$, $1 \leq t \leq \ell$, be a permutation of $\{1, 2, \ldots, q^r\}$. We have $\ell$ permutation, where the first permutation is the identity permutation. Let $\mathcal{C}_2$ be a perfect single-error-correcting Hamming code of length $\ell = \frac{q^{rs}-1}{q^r - 1}$ over an alphabet with $q^r$ letters. Let $\mathbb{C}_1^i$, $1 \leq i \leq q^r$, be the $i$th translate of $\mathbb{C}_1$, where $\mathbb{C}_1^1 = \mathbb{C}_1$. We construct the following code $\mathcal{C}^*$

$$\mathbb{C}^* = \{(x_{i_1}, \ldots, x_{i_\ell}) : x_{i_t} \in \mathbb{C}_1^{\pi_t(i_t)}, \ (i_1, \ldots, i_\ell) \in \mathcal{C}_2\}$$

*Theorem 14:* The code $\mathbb{C}^*$ is a perfect single-error-correcting Lee code of length $\frac{q^{rs}-1}{2}$ over an alphabet of size $\tau(2n+1)$.

*Proof:* Clearly, the length of the codewords from $\mathbb{C}^*$ is $\frac{q^r-1}{2}\frac{q^{rs}-1}{q^r-1} = \frac{q^{rs}-1}{2}$. The proof that the code $\mathbb{C}^*$ has minimum Lee distance three is identical to the related proof in Theorem 10.

By Corollary 1, the size of $\mathcal{C}_2$ is $\frac{q^{r\ell}}{1+(q^r-1)\ell} = q^{r\ell - rs}$. The size of $\mathbb{C}_1$ is $\tau^n(2n+1)^{n-1}$. Clearly, $|\mathbb{C}^*| = |\mathcal{C}_2| \cdot |\mathbb{C}_1|^\ell = q^{r\ell-rs}\tau^{n\ell}(2n+1)^{(n-1)\ell} = q^{r\ell-rs}\tau^{n\ell}q^{r(n-1)\ell} = q^{rn\ell - rs}\tau^{n\ell} = \frac{\tau^{n\ell}(2n+1)^{n\ell}}{q^{rs}}$.

This implies by Corollary 4 that $\mathbb{C}^*$ is a perfect single-error-correcting Lee code of length $\frac{q^{rs}-1}{2}$ over an alphabet of size $\tau(2n+1)$. ■

## V. ON THE NUMBER OF PERFECT CODES AND NON-PERIODIC CODES IN THE MANHATTAN METRIC

In this section we consider two problems related to perfect codes. The first one is the number of nonequivalent perfect codes, which was considered in many papers for the Hamming schemes, e.g. [22], [26]. It was proved that the number of nonequivalent perfect single-error-correcting code of length $n$ over GF($q$) is $q^{q^{cn}}$, where $c$ is a constant, $0 < c < 1$. The second problem is whether there exist non-periodic perfect and diameter perfect codes in $\mathbb{Z}^n$ with the Manhattan metric. Periodic and non-periodic codes were mentioned in [12], but for the best of our knowledge no construction of non-periodic codes was known until recently. We note that we ask the question on non-periodic codes in the Manhattan metric. The same question can be asked for the Lee metric; the answer seems to be even more complicated than the one for the Manhattan metric.

We will not go into the more complicated exact computations on the number of nonequivalent perfect and diameter perfect codes. We will just count the number of different perfect codes. The two product constructions given in Section IV can be used to provide a lower bound on the number of perfect codes and diameter perfect codes in the Lee metric. Given a diameter perfect code of length $p$, $p$ prime, the size of the related anticode is $4p$, and hence there are $2p$ even translates. Therefore, there are $(2p-1)!$ different perfect codes of length $2p$, given in the construction of Theorem 11. Continue iteratively with the construction of Theorem 11 we obtain

$$\prod_{i=1}^{r}(2^i p - 1)!^{2^{r-i}}$$

different diameter perfect codes of length $2^r p$ over $\mathbb{Z}_{4p}$.

Similarly, a bound on the number of perfect Lee code can be obtained from the construction of subsection IV-B. In this computation we can also take into account the bounds on the number of perfect codes in the Hamming scheme which are used in the construction.

Before we continue to discuss the topic of non-periodic perfect codes and diameter perfect codes in $\mathbb{Z}^n$ we consider a question related to both problems discussed in this section.

Let $C_1$ and $C_2$ be two distinct subcodes of $\mathcal{U}$, such that any elements in $\mathcal{U}$ is within Manhattan distance one from a unique codeword of $C_1$ and a unique codeword of $C_2$. If we consider anticodes instead of spheres then the elements of $C_1$ and $C_2$ are centers of the anticodes which form a tiling of $\mathcal{U}$. If $C_1$ is contained in a perfect single-error-correcting Manhattan code (or a diameter perfect code with minimum Manhattan distance four) $C$ then it can be replaced by $C_2$ to obtain a new different perfect single-error-correcting Manhattan code (or a diameter perfect code with minimum Manhattan distance four) $C'$. The technique was used in the Hamming scheme to form a large number of nonequivalent perfect codes [25], [26]. Same technique can be considered for the Lee metric.

The constructions in Section IV can provide such a space $\mathcal{U}$ and codes $C_1$ and $C_2$. If we change the permutation of the last coordinate by one transposition, it is easily verified the intersection of the two generated codes is relatively large (usually about $\frac{\eta-2}{\eta}$ of the code size, where $\eta = 2n$ for the construction of the diameter perfect codes and $\eta = q^r$ for the construction of the perfect single-error-correcting code). The two parts which are not in the intersection will have the role of $C_1$ and $C_2$. These can be the building block for non-periodic codes in the Manhattan metric. This is left as a problem for future research. Unfortunately, such a finite space $\mathcal{U} \subset \mathbb{Z}^n$ and codes $C_1$ and $C_2$ cannot exist. A subset $S$ is perfectly $\rho$-covered by $\mathbb{C}$ if for each element $s \in S$ there is a unique element $c \in \mathbb{C}$ such that $d(s,c) \leq \rho$.

*Theorem 15:* There is no finite subset of $\mathbb{Z}^n$ that can be perfectly $\rho$-covered by two different codes, using the Manhattan metric.

*Proof:* Assume the contrary; let $S$ be a subset of $\mathbb{Z}^n$ of smallest possible size, such that there exist two codes $C_1$ and $C_2$ which $\rho$-cover $S$. Let $k$ be the largest integer for which there exists a codeword in $C_1$ or $C_2$ with the value $k$ in one of the $n$ coordinates. W.l.o.g. we can assume that such a codeword is $(k, x_2, \ldots, x_n) \in C_1$. This codeword covers the word $(k + \rho, x_2, \ldots, x_n) \in S$. Since $k$ can be the largest integer in a codeword of $C_2$ then the only codeword of $C_2$ which can cover $(k + \rho, x_2, \ldots, x_n) \in S$ is $(k, x_2, \ldots, x_n)$. Thus, $C_1 \setminus \{(k, x_2, \ldots, x_n)\}$ and $C_2 \setminus \{(k, x_2, \ldots, x_n)\}$ perfectly $\rho$-cover a subset $S' \subset S$, a contradiction to the assumption that $S$ is such subset with the smallest size. Thus, there is no finite subset in $\mathbb{Z}^n$ that can be perfectly $\rho$-covered by two different codes. ∎

For each $n = 2^r$, $r \geq 1$, there exists a non-periodic diameter perfect code. If $n = 2$ then such a code exists for each even minimum Manhattan distance $2R + 2$, $1 \leq R$. We are using interleaving of the $R + 1$ lattices used to construct diameter perfect codes with minimum Manhattan distance $2R + 2$. For a given $R \geq 1$, let $S = \{s_i\}_{i=-\infty}^{\infty}$ be an infinite sequence, where $s_i \in \mathbb{Z}_{R+1}$. Given $S$ we construct the following set $\mathbb{T}$ of points

$$\mathbb{T} \stackrel{\text{def}}{=} \{(2(R+1)i + (R+1)j + s_i, (R+1)j + s_i) : s_i \in S, j \in \mathbb{Z}\}$$

The sequence $S$ will be called *non-periodic* if there are no nonzero integer $\rho$ and an integer $\tau$ such that $s_i = s_{i+\tau} + \rho \pmod{R+1}$.

*Theorem 16:* If the points of the set $\mathbb{T}$ are taken as centers for the spheres $S'_{2,R}$ then we obtain a tiling. The tiling is non-periodic if the sequence $S$ is non-periodic.

*Proof:* First we note that the set of centers $\{((R+1)j, (R+1)j) : j \in \mathbb{Z}\}$ forms a connected nonintersecting diagonal strip with the spheres $S'_{2,R}$. The same is true for the set of centers $\{(2(R+1)i + (R+1)j + s_i, (R+1)j + s_i) : j \in \mathbb{Z}\}$ for any given $i \in \mathbb{Z}$. The set of centers $\{(2(R+1)i + (R+1)j, (R+1)j) : i \in \mathbb{Z}, j \in \mathbb{Z}\}$ is exactly the lattice formed from the generator matrix

$$\begin{bmatrix} R+1 & R+1 \\ 2(R+1) & 0 \end{bmatrix},$$

which forms a diameter perfect code with minimum distance $2(R+1)$. Finally, replacing the set of centers $\{(2(R+1)i + (R+1)j, (R+1)j) : j \in \mathbb{Z}\}$ in this tiling by the set of centers $\{(2(R+1)i + (R+1)j + s_i, (R+1)j + s_i) : j \in \mathbb{Z}\}$ is just a shift of the diagonal strip in a 45 degrees diagonal direction. This does not affect the fact that the set of these centers forms a tiling with the spheres $S'_{2,R}$.

It is readily verified that if the sequence $S$ is non-periodic then also the tiling generated from $\mathbb{T}$ is non-periodic. ∎

The construction of two-dimensional non-periodic diameter perfect codes yields an uncountable number of diameter perfect codes. The number of nonequivalent perfect codes formed in this way is equal the number of infinite sequences over the alphabet $\mathbb{Z}_{R+1}$.

Finally, it is easy to verify the following theorem.

*Theorem 17:* Let $\mathbb{C}_1$ be a non-periodic diameter perfect code of length $n$ with minimum Manhattan distance four, and let $\mathbb{C}_2$ be a diameter perfect code of length $n$ and minimum Manhattan distance four. The code $\mathbb{C}^*$ defined in (4) is a non-periodic diameter perfect code of length $2n$ and minimum Manhattan distance four.

## VI. Conclusions and Open Problems

We have considered several questions regarding perfect codes in the Lee and Manhattan metrics. We gave two product constructions for perfect and diameter perfect codes in these metrics. One product construction is based on a new product construction for $q$-ary perfect codes in the Hamming scheme. the second product construction is based on the well known doubling construction for binary perfect codes in the Hamming scheme. We used our constructions to find a lower bound on the number of perfect and diameter perfect codes in the Lee and Manhattan metrics. We also used the constructions to prove the existence of non-periodic perfect diameter codes in $\mathbb{Z}^n$. Our discussion raises many open problem from which we choose a non-representative set of problems.

1) Find new product constructions for perfect Lee codes and diameter perfect lee codes with different parameters from those given in our discussion.
2) Prove that there are no nontrivial perfect codes, of length $n > 2$, with radius greater than one in the Manhattan metric.
3) Do there exist more diameter perfect codes with minimum distance greater than four in the Lee metric? Minkowski's lattice is the only known example.
4) For each $n > 4$, what is the smallest $m$ for which there exists $(n,4,4n,m)$ diameter perfect codes.
5) A tiling of $\mathbb{R}^n$ is called *regular* if neighboring anticodes meet along entire $(n-1)$-dimensional faces of the original cubes. Non-regular tiling of $\mathbb{R}^n$ exists if and only if $2n+1$ is not a prime [39]. Does there exist a non-regular tiling of $\mathbb{R}^n$ formed from anticodes with diameter 3?
6) Are there values of $n$ for which there are no two nonequivalent perfect single-error-correcting Lee codes of length $n$?
7) Find a construction for non-periodic perfect single-error-correcting Lee codes. Recently, such a construction was given in [40] for codes of length $n$, where $2n+1$ is not a prime, in the Manhattan metric.
8) Does there exists a non-periodic diameter perfect code with minimum Manhattan distance four for each length greater than one? What about the Lee metric?

### Acknowledgement

The author would like to devote this work for the memory of the late Rudolph Ahlswede with whom he had several interesting discussions during the last fifteen years on anticodes and perfect codes. During an ITW meeting in Ireland at the beginning of September 2010 he brought to his attention that the size of a maximum anticode in the Manhattan metric was found recently. This made it possible to call the codes in the Lee metric obtained in this paper, diameter perfect codes. The author would like to thank Peter Horak and Denis Krotov for their important comments and for providing him some important references.

## References

[1] C. Y. Lee, "Some properties of nonbinary error-correcting code", *IRE Trans. on Inform. Theory*, vol. 4, pp. 72–82, 1958.

[2] W. Ulrich, "Non-binary error correction codes", *Bell Sys. Journal*, vol. 36, pp. 1341–1387, 1957.

[3] S. W. Golomb and L. R. Welch, "Perfect codes in the Lee metric and the packing of polyominos", *SIAM Journal Applied Math.*, vol. 18, pp. 302–317, 1970.

[4] R. M. Roth and P. H. Siegel, "Lee-metric BCH codes and their application to constrained and partial-response channels", *IEEE Trans. on Inform. Theory*, vol. IT-40, pp. 1083–1096, July 1994.

[5] R. M. Roth, *Introduction to Coding Theory*, Cambridge University Press, 2005.

[6] S. Gravier, M. Mollard, and Ch. Payan, "On the existence of three-dimensional tiling in the Lee metric", *European J. Combin.*, vol. 19, pp. 567–572, 1998.

[7] S. Špacapan, "Non-existence of face-to-face four dimensional tiling in the Lee metric", *European J. Combin.*, vol. 28, pp. 127–133, 2007.

[8] P. Horak, "Tilings in Lee metric", *European J. Combin.*, vol. 30, pp. 480–489, 2009.

[9] K. A. Post, "Nonexistence theorem on perfect Lee codes over large alphabets", *Information and Control*, vol. 29, pp. 369–380, 1975.

[10] J. Astola, "On perfect Lee codes over small alphabets of odd cardinality", *Discrete Applied Mathematics*, vol. 4, pp. 227–228, 1982.

[11] P. Horak, "On perfect Lee codes", *Discrete Mathematics*, vol. 309, pp. 5551–5561, 2009.

[12] B. AlBdaiwi, P. Horak, and L. Milazzo, "Enumerating and decoding perfect linear Lee codes", *Designs, Codes Crypto.*, vol. 52, pp. 155-162, 2009.

[13] J. C.-Y. Chiang and J. K. Wolf, "On channels and codes for the Lee metric", *Information and Control*, vol. 19, pp. 1593–173, September 1971.

[14] C. Satyanarayana, "Lee metric codes over integer residue rings", *IEEE Trans. on Inform. Theory*, vol. IT-25, pp. 250–254, March 1979.

[15] K. Nakamura, "A class of error-correcting codes for DPSK channels", in *Proc. IEEE Conf. Commun.*, pp. 45.4.1–45.4.5, 1979.

[16] J. T. Astola, "Concatenated codes for the Lee metric", *IEEE Trans. on Inform. Theory*, vol. IT-28, pp. 778–779, September 1982.

[17] M. Blaum, J. Bruck, and A. Vardy, "Interleaving schemes for multi-dimensional cluster errors", *IEEE Trans. Inform. Theory*, vol. IT-44, pp. 730–743, March 1998.

[18] T. Etzion and E. Yaakobi, "Error-correction of multidimensional bursts", *IEEE Trans. on Inform. Theory*, vol. IT-55, pp. 961–976, March 2009.

[19] A. Barg and A. Mazumdar, "Codes in permutations and error correction for rank modulation", vol. IT-56, pp. 3158–3165, July 2010.

[20] K. T. Phelps, "A combinatorial construction of perfect codes", *SIAM J. Alg. Disc. Meth.*, vol. 4, pp. 398–403, 1983.

[21] K. T. Phelps, "A general product construction for error-correcting codes", *SIAM J. Alg. Disc. Meth.*, vol. 5, pp. 224–228, 1984.

[22] K. T. Phelps, "A product construction for perfect codes over arbitrary alphabets", *IEEE Trans. on Inform. Theory*, vol. IT-30, pp. 769–771, September 1984.

[23] M. Mollard, "A generalized parity function and its use in the construction of perfect codes", *SIAM J. Alg. Disc. Meth.*, vol. 7, pp. 113–115, 1986.

[24] V. A. Zinov'ev, "A combinatorial methods for the construction and analysis of nonlinear error-correcting codes", *Doc.. D. Thesis, Moscow*, 1988 (in Russian).

[25] T. Etzion and A. Vardy, "Perfect binary codes: constructions, properties, and enumeration", *IEEE Trans. on Inform. Theory*, vol. IT-40, pp. 754–763, 1994.

[26] T. Etzion, "Nonequivalent $q$-ary perfect codes", *SIAM Journal on Discrete Mathematics*, vol. 9, pp. 413–423, 1996.

[27] T. Etzion, "Tilings with generalized Lee spheres", in *Mathematical Properties of Sequences and other Combinatorial Structures*, editors J.-S. No, H.-Y. Song, T. Helleseth, and P. V. Kumar, pp. 181–198, 2002.

[28] T. Etzion, A. Vardy, and E. Yaakobi, "Dense error-correcting codes in the Lee metric", *Information Theory Workshop*, Dublin, August-September 2010.

[29] R. Ahlswede, H. K. Aydinian, and L. H. Khachatrian, "On perfect codes and related concepts", *Designs, Codes Crypto.*, vol. 22, pp. 221-237, 2001.

[30] G. A. Kabatyanskii and V. I. Panchenko, "Unit sphere packings and coverings of the Hamming space", *Prob. Inform. Transm*, vol. IT-34, pp. 3–16, 1988.

[31] H. O. Hämäläinen, "Two new binary codes with minimum distance three", *IEEE Trans. on Inform. Theory*, vol. IT-34, p. 875, July 1988.

[32] S. Litsyn and B. Mounits, "Improved lower bounds on the sizes of single-error correcting codes", *Designs, Codes Crypto.*, vol. 42, pp. 67–72, 2007.

[33] F. J. MacWilliams and N. J. A. Sloane, *The Theory of Error-Correcting Codes*, Amsterdam: North-Holland, 1977.

[34] M. Best, "Perfect codes hardly exist", *IEEE Trans. on Inform. Theory*, vol. IT-29, pp. 349–351, 1983.

[35] Ph. Delsarte, "An algebraic approach to association schemes of coding theory", *Philips J. Res.*, vol. 10, pp. 1–97, 1973.

[36] R. Ahlswede and V. Blinovsky, *Lectures on Advances in Combinatorics*, Springer-Verlag, 2008.

[37] H. Minkowski, "Dichteste gitterformige Lagerung kongruenter Korper", *Nachrichten Ges. Wiss. Gottingen*, pp. 311–355, 1904.

[38] D. S. Krotov, "$\mathbb{Z}_4$-linear Hadamard and extended perfect codes", *Electro. Notes Discrete Math.*, vol. 6, pp. 107–112, 2001.

[39] S. Szabó, "On mosaics consisting of multidimensional crosses", *Acta Math. Acad. Sci. hung.*, vol. 38, pp. 191–203, 1981.

[40] P. Horak and B. AlBdaiwi, "Non-periodic tilings of $\mathbb{R}^n$ by crosses", *Discrete and Computational Geometry*, submitted.